\documentstyle[12pt,epsfig]{article}
 \def\be{\begin{equation}}
 \def\ee{\end{equation}}
 \def\bea{\begin{eqnarray}}
 \def\eea{\end{eqnarray}}
 \def\nn{\nonumber}
 \def\vx{{\vec x}}
 \def\vk{{\vec k}}
 \def\vy{{\vec y}}
 \def\vz{{\vec z}}
 \def\p{\phi}
 \def\ppl{\phi_+}
 \def\pmi{\phi_-}

\begin{document}

 \title{$\phi^4$-Model and Holography in Four Dimensions}
 \author{Farhang Loran\thanks{e-mail:
loran@cc.iut.ac.ir}\\ \\
  {\it Department of  Physics, Isfahan University of Technology (IUT)}\\
{\it Isfahan,  Iran.} }

 \date{}
 \maketitle
 \begin{abstract}
 The generating function of correlators of dual operators on the boundary of $\mbox{(A)dS}_4$
 space corresponding to the conformally coupled $\phi^4$-model is obtained up to first order
 in the coupling
 constant by using the conformal map between massless scalar fields in (Euclidean) Minkowski
 space and conformally coupled scalars on (Euclidean anti) de Sitter space. Some exact
 classical solutions of the nonlinear wave equation of massless (conformally coupled)
 $\phi^3$, $\phi^4$ and $\phi^6$-models in $D=6,4,3$ Euclidean/Minkowski (AdS/dS)
 spaces are obtained.
 \end{abstract}

 \newpage
 \section{Introduction}\label{sec1}
 According to the generalized second law of thermodynamics \cite{Bek} the total entropy of
 ordinary matter and black hole will never decrease in any physical process. This idea
 has led to the holographic principle \cite{Hol,Bus} which is a relation between
 information and geometry and/or the entropy and area of an event horizon, $S\sim A$.
 According to the holographic principle the physics of a generic space-time can be described
 in terms of a theory on the boundary (the holographic screen). A quantitative realization
 of the holographic principle is provided by the AdS/CFT correspondence \cite{Mal}.
 The universality of the holographic principle and more specially the generalized second law
 of thermodynamics which includes de Sitter space \cite{Haw}
 as well as AdS space has motivated many attempts to obtain a holographic duality between
 physics in de sitter space and some CFT on its boundary. In the absence of a well-posed
 stringy description of dS/CFT correspondence different approaches to this problem has been
 considered \cite{Strom}. For example, holographic reduction of Minkowski space-time
 \cite{Sol} provides a duality between Minkowski space-time and a CFT defined on the boundary
 of the light cone. The key-point in that method is the fact that Minkowski space
 time can be sliced in terms of Euclidian AdS and Lorentzian dS
 slices which correspond to the time-like and space-like regions
 respectively. Therefore known facts about AdS/CFT duality can be used to
 understand Minkowski/CFT and dS/CFT correspondence.
 \par
 Recently in  \cite{map,Spin}, a holographic description of Euclidean/Minkowski is given
 for free massless scalar and Dirac fields. In that method, the action for
 free field theory on the boundary $t=0$ is derived by solving the action of
 free scalars/Dirac fields
 in terms of the Cauchy data on the hypersurface $t=0$. Using the conformal map between
 (Euclidean) Minkowski space and (Euclidean anti) de Sitter space which maps the
 hypersurface $t=0$ to the boundary of (A)dS space, it is  verified that the obtained
 action on
 the boundary $t=0$ is exactly the generating function of CFT correlators of the dual
 operators recognized in AdS/CFT correspondence \cite{Witten} and
 dS/CFT correspondence \cite{Strom,Les}.
 Here, by a similar method, we derive the generating function of correlators of
 boundary operators dual to the $\p^4$-model in $\mbox{(A)dS}_4$
 usnig a holographic description of $\p^4$-model in four dimensional (Euclidean)
 Minkowski space-time.
 \par
 The paper is organized as follows. In section 2, we study
 free massless scalar field theory on Euclidean space $R^{d+1}$ and the conformal map
 to Euclidean $\mbox{AdS}_{d+1}$. We show that the CFT action on the boundary of AdS space
 can be obtained by solving the action of free scalars on $R^{d+1}$ in terms of the Cauchy
 data at $t=0$. Finally we classify interacting scalar field theories that can be
 conformally mapped from $R^{d+1}$ to Euclidean $\mbox{AdS}_{d+1}$.
 In section 3, we study $D=4$ $\p^4$-model on Euclidean space.
 In section 4 some exact classical
 solutions of non-linear Klein-Gordon equation of massless (conformally coupled)
 $\phi^3$, $\phi^4$ and $\phi^6$-models in $D=6,4,3$ Euclidean/Minkowski (AdS/dS)
 spaces are obtained.
 Section 5 is devoted to scalar field theory on Minkowski space time and dS/CFT
 correspondence. We conclude in section 6. The free scalar field theory in
 curved spacetime is briefly reviewed in the appendix.
 \section{Massless Scalars on Euclidean Space}\label{sec2}
 The equation of motion for free massive scalar fields, or free
 conformally coupled scalars on Euclidean $\mbox{AdS}_{d+1}$ with metric,
 \be
  ds^2=\frac{1}{t^2}\left(dt^2+\sum_{i=1}^d dx_i^2\right),
 \label{met}\ee
 is
 \be
 \left(t^2\partial_t^2+(1-d)t\partial_t+t^2\nabla^2-m^2\right)\Phi=0,
 \label{a1}
 \ee
 where $\partial_t=\frac{\partial}{\partial t}$ and
 \be
 \nabla^2=\sum_{i=1}^d\frac{\partial^2}{\partial x_i^2}.
 \ee
 See the appendix for a brief review of scalar field theory in curved
 spacetimes.  One can show that if $\phi$ is a massless scalar
 field in $d+1$ dimensional Euclidean space $R^{d+1}$ with metric
 $ds^2=dt^2+dx_i^2,$ satisfying the equation
 \be
 (\partial_t^2+\nabla^2)\phi=0,
 \label{a2}
 \ee
 then
 \be
 \Phi=t^{\frac{d-1}{2}}\phi,
 \label{map}
 \ee
 is a solution of Eq.(\ref{a1}) with mass
 $m^2=\frac{1-d^2}{4}$. Since $\frac{-d^2}{4}<m^2<0$, this solution is
 stable in $AdS_{d+1}$.
 From AdS/CFT correspondence \cite{Witten} it is known that the dual theory on the
 boundary $t=0$ is a conformal theory with the following action,
 \be
 I[\phi]=\int d^dy d^dz
 \frac{\phi_0({\vec y})\phi_0({\vec z})}{\left|{\vec y}-{\vec
 z}\right|^{2(d+\lambda_+)}},
 \label{a3}
 \ee
 where $\lambda_+$ is the larger root of the equation $\lambda(\lambda+d)=m^2$.
 \footnote{$I[\phi]$
 is the generating function for CFT correlators i.e. $I[\phi]=\ln \left<\exp \int \p
 {\cal O}\right>$ where ${\cal O}$ is a dual operator. } Here
 $\lambda_+=\frac{(1-d)}{2}$ as far as $m^2=\frac{(1-d^2)}{4}$.
 $\phi_0$ is a function on the boundary such that $\Phi(\vx,t)\sim t^{-\lambda_+}\phi_0$
 as $t\to 0$. From the map (\ref{map}), one can interpret $\phi_0(\vx)$ in
 Euclidean space  $R^{d+1}$ as the initial data on the hypersurface $t=0$.
 Therefore one expects that the action (\ref{a3}) can be obtained from the action of scalar
 fields in $d+1$ dimensional Euclidean space,
 \be
 I[\phi]=\frac{1}{2}\int
 dtd^dx\left((\partial_t\phi)^2+(\nabla\phi)^2\right),
 \label{EA}
 \ee
 if one solves equation (\ref{a2})
 in terms of the initial data $\phi_0(\vx)$ given on the
 hypersurface $t=0$. Proof is as follows:
 \par
 The most general solution of the equation of motion that vanishes as $t$
 tends to infinity is
 \be
 \phi(\vx,t)=\int d^dk {\tilde \phi}(\vk)e^{i\vk.\vx}e^{-\omega
 t},
 \label{a4}
 \ee
 where $\omega=\left|\vk\right|$ and
 \be
 {\tilde \phi}(\vk)=\int d^dx \phi_0(\vx)e^{-i\vk.\vx}.
 \label {a41}
 \ee
 Here, $\p_0(\vx)$ is the initial value on the boundary
 $t=0 $. Inserting (\ref{a41}) into (\ref{a4}) one obtains
 \be
 \p(\vx,t)=\int d^dy\ {\cal G}(\vx,t;\vy)\p_0(\vy),
 \label{a42}
 \ee
 where
 \be
 {\cal G}(\vx,t;\vy)=\int d^dk\  e^{-\omega t} e^{i\vk.(\vx-\vy)}.
 \label{g}
 \ee
 ${\cal G}(\vx,t;\vy)$ is the solution of wave equation i.e. $\Box {\cal G}=0$,
 with the initial condition ${\cal G}(\vx,0;\vy)=\delta^d(\vx-\vy)$.
 To obtain the action of the corresponding theory on the boundary $t=0$ one should insert
 (\ref{a42}) into (\ref{EA}). But it is more suitable to rewrite the action (\ref{EA}) in
 the form,
 \be
 I[\phi]=-\frac{1}{2}\int d^{d+1}x\ \p\Box\p-\frac{1}{2}\int d^dx\
 \p_0(\vx)\partial_t\p_0(\vx),
 \label{intbypart}
 \ee
 which is obtained by an integration by part and under the assumption that
 $\phi(x)$ vanishes as $t$ tends to infinity and also at spatial
 infinity. Inserting (\ref{a42}) into (\ref{intbypart}), the first term
 vanishes and from the second term one obtains:
 \be
 I[\phi]=\frac{1}{2}\int d^dyd^dz\ \phi_0({\vec y})\phi_0({\vec z})F({\vec y}-{\vec
 z}),
 \label{a5}
 \ee
 in which
 \be
 F(\vx)=\int d^dk\  \omega e^{i\vk.\vx}.
 \label{a5-1}
 \ee
 As can be verified from the rotational invariance ($\vx\to {\bf R}\vx$,
 ${\bf R}\in SO(d)$),
 $F(\vx)$ depends only on the norm of $\vx$. By scaling $\vx$ by a factor $\lambda>0$ one
 can also show that $F(\lambda\vx)=\lambda^{-(d+1)}F(\vx)$ and consequently,
 \be
 \vx.\nabla F(\vx)=\lim_{\lambda\to 1}\frac{F(\lambda\left|\vx\right|)-F(\left|\vx\right|)}
 {\lambda-1}=-(d+1)F(\left|\vx\right|).
 \ee
 Therefore, $F(\vx)=\mbox{Const.}\left|\vx\right|^{-(d+1)}$ and
 \be
 I[\phi]=\mbox{Const.}\int d^dyd^dz\ \frac{\phi_0({\vec y})\phi_0({\vec
 z})}{ \left|{\vec y}-{\vec z}\right|^{d+1}},
 \label{af}
 \ee
 which is equal to (\ref{a3}) derived in AdS/CFT correspondence.
 \par
  One should note that in addition to the general solution
 (\ref{a4}), Eq.(\ref{a2}) has one further solution $\phi(\vx,t)=\alpha
 t$ where $\alpha$ is some constant that can not be determined from
 the initial data
 $\phi_0(\vx)$. Although this solution is meaningless in Euclidean
 space, but $\alpha$ is equal to the value of $\Phi(\vx,t)$ at $t\to \infty$ which is an
 additional point on the boundary of AdS.
 \par
 One can interpret the above result as a holographic
 description of Euclidean space. But our analysis considers only
 massless scalars. As can be easily verified, it is not possible to generalize the map
 (\ref{map}) to include massive scalar fields in $R^{d+1}$. A reason
 for this is the fact that the equation of motion of scalar fields
 $(\Box +m^2)\phi=0$ is not conformally covariant unless $m=0$.
 \par
 By the same consideration, one can also show that the map between free scalar field
 theories can only be generalized to $D=6$ $\p^3$-model, $D=4$ $\p^4$-model and $D=3$
 $\p^6$-model in which no specific length (mass) is given in the model as far as
 $g$, the coupling, is dimensionless. The proof is as follows: using the identity,
 \be
 \left(t^2\partial_t^2+(1-d)t\partial_t-t^2\nabla^2+\frac{d^2-1}{4}\right)\Phi(\vx,t)=
 t^{\frac{d-1}{2}+2}\left(\partial_t^2-\nabla^2\right)\phi,
 \ee
 in which $\Phi=t^{\frac{d-1}{2}}\phi$, and considering the potential
 $V(\phi)=\frac{g}{n+1}\phi^{n+1}$, one can show that
 \be
 n=1+\frac{4}{d-1},
 \label{n}\ee
 if $ \left(\partial_t^2-\nabla^2\right)\phi=-g\p^n,$
 and
 \be
 \left(t^2\partial_t^2+(1-d)t\partial_t-t^2\nabla^2+\frac{d^2-1}{4}\right)\Phi(\vx,t)=
 -g\Phi^n.
 \ee
 Since $n$ is an integer, the identity (\ref{n}) can be
 satisfied only for $d=2,3,5$ and $n=5,3,2$ respectively. All such theories
 are renormalizable.
 \section{$D=4$ Massless $\phi^4$-Model}\label{sec3}
 In this section we study $D=4$ massless $\phi^4$-model and obtain
 the corresponding action on the boundary $t=0$ up to first order in
 $g$, the coupling constant. As is shown in section~\ref{sec2}, this massless
 $\p^4$ model can be
 directly mapped to conformally coupled scalar theory on Euclidean AdS.
 The same method can be used to obtained the action on the boundary corresponding to
 $D=6$ $\p^3$-model and $D=3$ $\p^6$-model.
 \par
 The action is
 \bea
 I_g[\phi]&=&\int d^4x\left(\frac{1}{2}\left((\partial_t\phi)^2+(\nabla\phi)^2\right)-
 \frac{g}{4}\p^4\right),\nn\\
 &=&-\int_{R^4}
 \left(\frac{1}{2}\p\Box\p+\frac{g}{4}\p^4\right)-\frac{1}{2}\int_{R^3}
 \p_0(\vx)\partial_t\p_0(\vx).
 \label{b1}
 \eea
 The corresponding Eule-Lagrange equation of motion is the a non-linear Klein-Gordon equation
 on $R^4$ given as follows:
 \be
 \Box \p+g\p^3=0.
 \label{b2}
 \ee
 As before, to obtain the second equality in Eq.(\ref{b1}), an integration by part is
 made. It is also assumed that $\p(x)$ vanishes as $t\to\infty$ and
 at spatial infinity.  $\p_0(\vx)$  is the Cauchy data at $t=0$.
 In the next section we obtain some exact solutions of
 Eq.(\ref{b2}), but here we are interested in solutions only correct up to ${\cal O}(g^2)$.
 A formal solution of Eq.(\ref{b2}) is
 \be
 \p(x)=\eta(x)-g\int d^4y\ G_E(x;y)\p^3(y),
 \label{b3}
 \ee
 where $\Box \eta=0$ and $\Box G_E(x;y)=\delta_E^4(x-y)$.
 $G_E$ is the Euclidean Green function that is equal to the the Green function,
 \be
 G(x,y)\sim \int d^3k
 \frac{e^{i\omega(x^0-y^0)}}{\omega}e^{i\vk.(\vx-\vy)},\hspace{1cm}
 \omega=\left|\vk\right|,
 \label{Green}
 \ee
 after a Wick rotation $x^0\to i x^0$. Using ${\cal G}$
 defined in Eq.(\ref{g}), one can solve Eq.(\ref{b3}) in terms of $\p_0(\vx)=\p(x)|_{t=0}$
 as follows:
 \be
 \p(x)=\eta_{(1)}(x)-g\int d^4y\ G(x;y)\eta_{(0)}^3(y)+{\cal O}(g^2),
 \label{b4}
 \ee
 where $\eta_{0}(x)=\int d^3y\ {\cal G}(\vx,0;\vy)\p_0(\vy)$ and
 \be
 \eta_{(1)}(x)=\eta_{(0)}(x)+g\int d^4y \int d^3z\ {\cal
 G}(\vx,t;\vz)G(\vz,0;y)\eta_{(0)}^3(y).
 \label{b5}
 \ee
 One can verify that $\Box\eta_{(1)}=\Box\eta_{(0)}=0$ and
 $\eta_{(0)}(\vx,0)=\p_0(\vx)$.
 Since,
 \be
 \int d^3z\ {\cal G}(\vx,t;\vz)G(\vz,0;y)=G(x;y).
 \label{G-G}
 \ee
 as can be verified from Eqs.(\ref{g}) and (\ref{Green}) and
 using equations (\ref{b4}) and (\ref{b5}) one obtains,
 \be
 \p(x)=\eta_{(0)}(x)+{\cal O}(g^2).
 \label{important}
 \ee
 Consequently,
 \bea
 I_g[\p]&=&\frac{1}{2}\int d^4x \partial_\mu\eta_0\partial^\mu\eta_0-\frac{g}{4}\int
 d^4x\ \eta_0^4\nn\\
 &=&\mbox{Const.}\int d^dyd^dz\ \frac{\phi_0({\vec y})\phi_0({\vec z})}
 {\left|{\vec y}-{\vec z}\right|^4}-\frac{g}{4}\int d^4x\ \eta_0^4,
 \label{b6}
 \eea
 where to obtain the second equality we have used Eq.(\ref{af}).
 Consequently the only $g$-contribution in $I_g[\p]$ is from the
 term $\int_{R^4}\eta_{(0)}^4(x)$ that can be written as follows:
 \be
 \int_{R^4}\eta_{(0)}^4(x)=\int_{R^3}K_E(\vx_1,\cdots,\vx_4)\p_0(\vx_1)\cdots\p_0(\vx_4).
 \label{b10}
 \ee
 \bea
 K_E(\vx_1,\cdots,\vx_4)&=&\int d^4x\ {\cal G}(x;\vx_1)\cdots{\cal
 G}(x;\vx_4)\nn\\
 &=&\int d^3k_1\cdots d^3k_4\frac{e^{i\vk_1.\vx_1}\cdots
 e^{i\vk_4.\vx_4}}{\omega_1+\cdots+\omega_4}\delta^3(\vk_1+\cdots+\vk_4),
 \label{b11}
 \eea
 where to obtain the second equality we have used definition
 (\ref{g}). To calculate $K_E(\vx_1,\cdots,\vx_4)$ we note that
 \begin{enumerate}
 \item{$K_E$ is symmetric:
 $K_E(\vx_1,\cdots,\vx_4)=K_E(\vx_{p_1},\cdots,\vx_{p_4})$ for any
 permutation $p$,}
 \item{$K_E$ is invariant under translation: $K_E(\vx_1+{\vec a},\cdots,\vx_4+{\vec a})=
 K_E(\vx_1,\cdots,\vx_4)$,}
 \item{$K_E(\lambda\vx_1,\cdots,\lambda\vx_4)=\lambda^{-8}K_E(\vx_1,\cdots,\vx_4)$,
 for any real-valued positive $\lambda$,}
 \item{$\int d^3x_4\ K_E(\vx_1,\cdots,\vx_4)=K_E(\vx_1,\vx_2,\vx_3),$}
 \end{enumerate}
 where
 \be
 K_E(\vx_1,\vx_2,\vx_3)=\int d^3k_1d^3k_2d^3k_3
 \frac{e^{i(\vk_1.\vx_1+\vk_2.\vx_2+\vk_3.\vx_3)}}
 {\omega_1+\omega_2+\omega_3}\delta^3(\vk_1+\vk_2+\vk_3).
 \label{b12}
 \ee
 $K_E(\vx_1,\vx_2,\vx_3)$ is symmetric and invariant under translation,
 $K_E(\lambda\vx_1,\lambda\vx_2,\lambda\vx_3)=\lambda^{-5}K_E(\vx_1,\vx_2,\vx_3)$
 and
 \bea
 \int d^3x_3\ K_E(\vx_1,\vx_2,\vx_3)&=&\int d^3k_1d^3k_2\frac{e^{i(\vk_1.\vx_1+\vk_2.\vx_2)}}
 {\omega_1+\omega_2}\delta^3(\vk_1+\vk_2)\nn\\
 &=& \mbox{Const.}\frac{1}{\left|\vx_1-\vx_2\right|^2}.
 \label{b13}
 \eea
 From the above observations we conclude that
 \be
 K_E(\vx_1,\vx_2,\vx_3)=\mbox{Const.}\left(
 \left|\vx_1-\vx_2\right|\left|\vx_2-\vx_3\right|\left|\vx_3-\vx_1\right|\right)
 ^{-\frac{5}{3}},
 \label{b14}
 \ee
 as far as
 \be
 M(\vx_1,\vx_2)=\int d^3\vx_3 \
 (\left|\vx_2-\vx_3\right|\left|\vx_3-\vx_1\right|)^\frac{-5}{3}\sim
 \left|\vx_1-\vx_2\right|^\frac{-1}{3},
 \label{b15}
 \ee
 and consequently Eq.(\ref{b14}) is satisfied. The validity of
 Eq.(\ref{b15}) can be verified by noting  that
 $M(\vx_1,\vx_2)=M(\left|\vx_1-\vx_2\right|)$ and
 $M(\lambda\vx_1,\lambda\vx_2)=\lambda^\frac{-1}{3}M(\vx_1,\vx_2)$.
 Using this result we claim that,
 \be
 K_E(\vx_1,\cdots,\vx_4)=\mbox{Const.}(\left|\vx_1-\vx_2\right|\left|\vx_1-\vx_3\right|
 \left|\vx_1-\vx_4\right|\left|\vx_2-\vx_3\right|\left|\vx_2-\vx_4\right|
 \left|\vx_3-\vx_4\right|)^{-\frac{8}{6}}.
 \label{claim}\ee
 Summarizing our results, we have found that
 \bea
 I_g[\phi]&=&\int_{R^4}\left(\frac{1}{2}\left((\partial_t\phi)^2+(\nabla\phi)^2\right)-
 \frac{g}{4}\p^4\right)\nn\\&\sim&\mbox{Const.}\int_{R^3}\frac{\phi_0({\vec y})\phi_0({\vec
 z})}{ \left|{\vec y}-{\vec z}\right|^4}\nn\\
 &+&g\int_{R^3}\frac{\p_0(\vx_1)\p_0(\vx_2)\p_0(\vx_3)\p_0(\vx_4)}
 {(\left|\vx_1-\vx_2\right|\left|\vx_1-\vx_3\right|
 \left|\vx_1-\vx_4\right|\left|\vx_2-\vx_3\right|\left|\vx_2-\vx_4\right|
 \left|\vx_3-\vx_4\right|)^\frac{8}{6}}.
 \eea
 \section{Exact Solutions of Non-Linear Wave Equation}\label{sec4}
 In sections \ref{sec2} and \ref{sec3} we inserted the
 solutions of the Klein-Gordon equation into the action in order to
 find an action for the fields on the boundary $t=0$. The solutions of Euler-Lagrange
 equations obtained from the action on the boundary are the only configurations that can be
 considered as initial condition (Cauchy-data) which solve the non-linear wave
 equation (Klein-Gordon equation) with solutions that vanish as $t$ tends to
 infinity, since they are {\em on-shell} configurations by construction. In
 section~\ref{sec2} we verified that the action on the boundary ($R^{d}$) that was
 obtained by inserting the exact solutions of the mass-less
 free field equation into the action of fields in the bulk $R^{d+1}$
 is exactly the CFT action recognized in $\mbox{AdS}_{d+1}/\mbox{CFT}_d$ correspondence. In
 section~\ref{sec3} we considered $\p^4$-model in four dimension
 ($R^4$) and derived the action on the boundary $R^3$ (at $t=0$) by
 solving the action in terms of solutions of the non-linear wave equation obtained by
 perturbation. As far as the massless $\p^4$-model in $R^4$ is in
 one-to-one correspondence to conformally coupled $\p^4$ scalar theory on Euclidean AdS
 space, one may conjecture that the corresponding (CFT) action on the
 boundary of AdS space can be exactly obtained if one repeats the steps of
 sections \ref{sec2} and \ref{sec3} using the exact solutions of the non-linear Klein-Gordon
 equation. In what follows we consider $\p^n$-model in $d+1$ dimension and
 derive the exact plane wave solutions. Furthermore, in the case of
 conformally coupled scalar theories we also obtain some $SO(d+1)$-invariant
 solutions. As we will see for all such solutions, the scalar field is determined
 at all space-time points even on the boundary and no initial condition
 in the sense of free field solutions or solutions obtained by perturbation exist. The
 initial condition in the case of non-linear Klein-Gordon equation
 can at most clarify which of the solutions among the others should be considered
 and determine some constant coefficients that may appear in such
 solutions. By inserting any of the exact solutions into the
 action one does not obtain an action governing the fields on the boundary.
 This incommodity is caused by the lack of any superposition rule for solutions of
 non-linear differential equations. As far as the exact classical solutions can not be
 quantized by the same reason and one should use perturbation to obtain the $\p^4$
 {\em quantum field theory}, perturbation can also be considered as a reasonable method
 to obtain the boundary action.
 \subsection*{Plane Wave Solutions}
 The non-linear Klein-Gordon equation in $d+1$ dimension for plane
 wave solutions is,
 \be
 \frac{d^2}{du^2}\p(u)+\frac{1}{k^2}\frac{d V(\p)}{d\p}=0,
 \label{c1}
 \ee
 where $u=k_\mu x^\mu$ and $k^2=k_\mu k^\mu$ for some wave vector
 $k=(k_1,\cdots,k_{d+1})$. To obtain Eq.(\ref{c1}) we have inserted
 $\Box\p(u)=k^2\frac{d^2}{du^2}\p(u)$ into the Klein-Gordon equation
 $\Box \p+\frac{d V(\p)}{d\p}=0$. From Eq.(\ref{c1}) one verifies that plane wave solutions
 are not sensitive to the dimension of space-time.  The most general solutions of
 Eq.(\ref{c1}) are a one-parameter family that can be obtained by solving the following
 integral equation:
 \be
 \int \frac{d\p}{\sqrt{\frac{-2}{k^2}V(\p)+c}}=u,
 \label{c2}
 \ee
 where $c$ is some real-valued constant. For example if
 $V(\p)=\frac{g}{4}\p^4$, ($g<0$) then,
 \be
 \p_{c=0}(x)=\frac{1}{{\tilde k}.x},\hspace{1cm}\ {\tilde
 k}^2=-\frac{g}{2}\hspace{5mm}\left({\tilde
 k}^\mu=\sqrt{\frac{-g}{2}}\frac{k^\mu}{\sqrt{k^2}}\right).
 \ee
 \subsection*{$SO(d+1)$-Invariant Solutions}
 The Klein-Gordon equation for $SO(d+1)$-invariant solutions $\p=\p(s)$, where
 $s=\sqrt{x_\mu x^\mu}$ is
 \be
 \left(\frac{d^2}{ds^2}+\frac{d}{s}\frac{d}{d s}\right)\p+g\p^n=0.
 \label{d1}
 \ee
 One solution of this equation is
 \be
 \p_0(s)=\left(\frac{2(d-1-\frac{2}{n-1})}{(n-1)}\right)^\frac{1}{n-1}
 \left(\frac{1}{g s^2}\right)^\frac{1}{n-1}
 \label{d2}
 \ee
 These solutions can be obtained by considering the ansatz  $\p(s)=\alpha
 s^\beta$ and solving the wave equation to determine $\alpha$ and $\beta$.
 The above solutions become singular as $g\to 0$. One can show that
 Eq.(\ref{d1}) has also solutions like,
 \be
 \p(s)=\frac{\alpha}{(\beta+s^2)^\gamma},
 \label{d3}
 \ee
 only for conformally coupled theories, i.e. for $d=2,3,5$ and $n=5,3,2$
 respectively. The solutions are:
 \bea
 \frac{\alpha}{\left(\frac{\alpha}{24}g+s^2\right)^2},&\hspace{5mm}&n=2,\ d=5,\nn\\
 \frac{\alpha}{\left(\frac{\alpha^2}{8}g+s^2\right)},&&n=3,\ d=3,\nn\\
 \frac{\alpha}{\sqrt{\frac{\alpha^4}{3}g+s^2}},&&n=5,\ d=2,
 \label{d4}
 \eea
 where $\alpha$ is some arbitrary real-valued constant. By Wick
 rotation $t\to it$ one obtains the solutions of wave equation
 in Minkowski space-time. Using the map
 $\p\to\Phi=t^\frac{d-1}{2}\p$,
 defined  in section~\ref{sec2} one can also obtain the corresponding
 $SO(d)$-invariant solutions in Euclidean $\mbox{AdS}_{d+1}$ space and
 $\mbox{dS}_{d+1}$ space. For example, a solution for the
 Klein-Gordon equation for $\p^4$ model in the ${\cal O}^-$ region of
 $\mbox{dS}_4$ space is
 \be
 \p^-_{\mbox{dS}_4}(t,\vx)=\frac{\alpha
 t}{\left(\frac{\alpha^2}{8}g-t^2+\left|\vx\right|^2\right)},\hspace{1cm}\alpha\in R.
 \label{d5}
 \ee
 A method to obtain solutions given in Eq.(\ref{d4}) and more such
 solutions for conformally coupled models is as follows. Using the
 solutions $\p_0(s)$ given in Eq.(\ref{d2}) one can try to solve the Klein-Gordon
 equation for $\p(s)=\p_0(s)\eta(s)$. The resulting equation for
 $\eta(s)$ is
 \be
 \left(s\frac{d}{d s}\right)\left(s\frac{d}{d s}\right)\eta+
 \left\{\begin{array}{cc}4\eta(\eta-1)=0,&
 n=2, \ d=5,\\
 \eta(\eta^2-1)=0,& n=3,\ d=3,\\
 \frac{1}{4}\eta(\eta^4-1)=0,&n=5,\ d=2.
 \end{array}\right.
 \label{d6}
 \ee
 The solutions given in Eq.(\ref{d4}) are obtained by solving
 Eq.(\ref{d6}) with vanishing constant of integration.
 \section{Scalars on Minkowski Space}\label{sec5}
 In this section we consider $\p^4$ scalar field theory on Minkowski
 space time. We solve the Klein-Gordon equation in terms of the
 initial conditions $\ppl(\vx)=\p(\vx,0)$ and $\pmi(\vx)=i\partial_t\p(\vx,0)$
 up to first order in $g$
 and obtain the action on the boundary by solving the $D=4$ $\p^4$-action
 in terms of these solutions. In the case of free field theory, we
 show that this action is similar to the boundary action found in
 dS/CFT for conformally coupled scalars.
 \subsection{Free-Scalar Theory and dS/CFT
 Correspondence}\label{sec51}
  Similar to section \ref{sec2}, one can show that massless scalar fields in
 $d+1$ dimensional Minkowski space-time $M_{d+1}$ can be mapped
 by (\ref{map}) to scalars with mass $m^2=\frac{d^2-1}{4}$  or the
 conformally coupled scalars on $dS_{d+1}$. To verify this, one can use the following
 metrics for $dS_{d+1}$ and $M_{d+1}$ respectively:
 \be
 ds^2_{dS}=\frac{1}{t^2}\left(-dt^2+\sum_{i=1}^d
 dx_i^2\right),
 \label{e1}
 \ee
 \be
 ds^2_{M}=\left(-dt^2+\sum_{i=1}^d
 dx_i^2\right)
 \label{e2}
 \ee
 The metric (\ref{e1}) covers only half of dS space. This region called ${\cal O}^-$ is the
 region observed by an observer on the south pole ${\cal I}^-$
 but is behind the horizon of the
 observer on the north pole ${\cal I}^+$. By construction  $t>0$.
 Following the Strominger proposal \cite{Strom}, dual operators living on the
 boundary ${\cal I}^-$ obey,
 \be
 \left<{\cal O}_\phi(z,{\bar z}),{\cal O}_\phi(v,{\bar
 v})\right>=\frac{\mbox{const.}}{\left|z-v\right|^{2h_+}},
 \label{e3}
 \ee
 where
 \be
 h_\pm=\frac{1}{2}\left(d\pm\sqrt{d^2-4m^2}\right).
 \label{e4}
 \ee
 Again, the existence of the map (\ref{map}) suggests that
 Eq.(\ref{e3}) can be obtained by solving the equations of motion of massless scalar
 fields on $M_{d+1}$ in terms of the initial data at $t=0$. If yes
 then the final result can be interpreted  as a holographic
 description of massless scalars in Minkowski space time. Such a
 description can be made covariant  by considering a covariant boundary \cite{Bus}
 instead of the hypersurface $t=0$, which here corresponds to the ${\cal I}^-$ by
 (\ref{map}).
 \par
 General arguments show that a massive scalar field
 behaves as $t^{h_\pm}\phi_\pm$ near ${\cal I}^-$ \cite{Strom}. Since $h_-=\frac{d-1}{2}$ for
 $m^2=\frac{d^2-1}{4}$, using Eq.(\ref{map}) one verifies that,
 $\phi_-(\vx)=\phi(\vx,t)|_{t=0}$. As will be exactly
 shown, $\phi_+(\vx)=i\partial_t\phi(\vx,t)|_{t=0}$.
 Two evidences for this claim are:
 \begin{enumerate}
 \item{Since $h_+=h_-+1$ for $m^2=\frac{d^2-1}{4}$, using Eq.(\ref{map})
 one can verify that $\partial_t\phi$ mapped to $dS_{d+1}$
 behaves as $t^{h_+}$ near ${\cal I}^-$ as demanded.}
 \item{A general solution of equation of motion for massless scalars in $M_{d+1}$,
 contains oscillating terms with both positive and
 negative frequencies. Therefore $\phi(\vx,t)$ can only be given in terms of both
 $\phi_0(\vx,0)$ and $\partial_t\phi_0(\vx,0)$.}
 \end{enumerate}
 \par
 In other words the main purpose of this subsection is to derive the
 CFT  action of the dual theory on the boundary of dS space,
 \be
 I[\phi]=\int_{{\cal I}^-}d^dyd^d(z)\left(\frac{\phi_-({\vec
 y})\phi_-({\vec z})}{\left|{\vec y}-{\vec z}\right|^{2h_+}}+
 \frac{\phi_+({\vec
 y})\phi_+({\vec z})}{\left|{\vec y}-{\vec
 z}\right|^{2h_-}}\right)
 \label{int1}
 \ee
 obtained in ds/CFT correspondence, by inserting the solution of Klein-Gordon
 equation for massless scalar fields in $M_{d+1}$,
 \be
 \left(\partial_t^2-\nabla^2\right)\phi=0,
 \label{KG}
 \ee
 satisfying the initial conditions
 \be
 \phi(\vx,0)=\phi_-(\vx),
 \hspace{1cm}\partial_t\phi(\vx,0)=i\phi_+(\vx),
 \label{IC}
 \ee
 into the action of massless scalar fields in $M_{d+1}$,
 \be
 I[\phi]=\frac{1}{2}\int
 dtd^dx\left((\partial_t\phi)^2-(\nabla\phi)^2\right).
 \label{MA}
 \ee
 The most general solution of Eq.(\ref{KG}) satisfying the initial conditions (\ref{IC}) is
 \bea
 \phi(\vx,t)&=&\frac{1}{2}\int d^dyd^dk
 \left(\phi_-(\vy)-\frac{\phi_+(\vy)}{\omega}\right)
 e^{i\vk.(\vx-\vy)}e^{-i\omega t}\nn\\
 &+& \frac{1}{2}\int d^dyd^dk \left(\phi_-(\vy)+\frac{\phi_+(\vy)}{\omega}\right)
 e^{i\vk.(\vx-\vy)}e^{i\omega t}\nn\\
 &=&\int d^d y \left({\cal G}_-(x;\vy)\pmi(\vy)+{\cal
 G}_+(x;\vy)\ppl(\vy)\right),
 \label{e5}
 \eea
 where $\omega=\left|\vk\right|$ and
 \bea
 {\cal G}_-(\vx,t;\vy)&=&\int d^dk
 e^{i\vk.(\vx-\vy)}\cos(\omega t),\nn\\
 {\cal G}_+(\vx,t;\vy)&=&i\int d^dk
 e^{i\vk.(\vx-\vy)}\frac{\sin(\omega t)}{\omega}.
 \label{e6}
 \eea
 To obtain the action (\ref{MA}) in terms of the solution (\ref{e5}), one
 should do calculations like:
 \bea
 \int dtd^dx\partial_t{\cal G}_-(\vx,t;\vy)\partial_t{\cal
 G}_-(\vx,t;\vz)&=&\int d^d k\ \omega^2 e^{i\vk.(\vy-\vz)}\int dt\sin^2(\omega t)\nn\\
 &=&\int d^dk\ \omega^2e^{i\vk.(\vy-\vz)}\left(\frac{\pi}{\omega}\right)\nn\\
 &\sim& \frac{1}{\left|\vy-\vz\right|^{d+1}}.
 \label{e7}
 \eea
 and
 \be
 \int dtd^dx\partial_t{\cal G}_-(\vx,t;\vy)\partial_t{\cal
 G}_+(\vx,t;\vz)=i\int d^d k\ \omega e^{i\vk.(\vy-\vz)}\int dt\sin(\omega t)\cos(\omega
 t)=0
 \label{e8}
 \ee
 Finally one obtains,
 \be
 I[\p]\sim\int d^dy d^dz\ \left(\frac{\phi_-(\vy)\phi_-(\vz)}{\left|\vz-\vy\right|^{d+1}}+
 \frac{\phi_+(\vy)\phi_+(\vz)}{\left|\vz-\vy\right|^{d-1}}\right).
 \label{e9}
 \ee
 \subsection{Massless $\phi^4$-Model on $M_4$}\label{52}
 In this subsection we repeat the calculations of section \ref{sec3} in the case
 of Minkowski space-time. As is explained before, the boundary action that we obtain here
 is related to the action on the boundary of $\mbox{dS}_4$ space corresponding to the
 conformally coupled $\p^4$-model.
 \par
 The solution of the Klein-Gordon equation $\Box\p+g\p^3=0$ is
 \be
 \p(x)=\eta_{(1)}(x)-g\int d^4y\ G(x;y)\eta_{(0)}^3(y)+{\cal O}(g^2),
 \label{ee1}
 \ee
 where
 \be
 \eta_{(0)}(x)=\int d^3 y \left({\cal G}_-(x;\vy)\pmi(\vy)+{\cal
 G}_+(x;\vy)\ppl(\vy)\right),
 \label{ee2}
 \ee
 and
 \be
 \eta_{(1)}(x)=\eta_{(0)}(x)+g\int d^4y \left(\int d^3z {\cal
 G}_-(x;\vz)G(\vz,0;y)+i{\cal G}_2(\vz,0;y)\partial_t
 G(\vz,0;y)\right)\eta^3_{(0)}(y).
 \label{ee3}
 \ee
 Similar to Eq.(\ref{important}), it is easy to verify that
 $\p(x)=\eta_{(0)}(x)+{\cal O}(g^2)$ and consequently the action is,
 \bea
 I_g[\p]&=&\int_{M_4}
 \left(\frac{1}{2}\partial_\mu\p\partial^\mu\p-\frac{g}{4}\p^4\right)\nn\\
&=&\mbox{Const.}\int_{R^3} \
\left(\frac{\phi_-(\vy)\phi_-(\vz)}{\left|\vz-\vy\right|^4}+
 \frac{\phi_+(\vy)\phi_+(\vz)}{\left|\vz-\vy\right|^2}\right)\nn\\
 &-&\frac{g}{4}\int_{R^3}K_-(\vx_1,\cdots,\vx_4)\pmi(\vx_1)\cdots\pmi(\vx_4)+
 \ \left(-\to +\right) \nn\\
 &-&\frac{g}{4}\int_{R^3}{\tilde K}(\vx_1,\vx_2,\vx_3,\vx_4)\pmi(\vx_1)\pmi(\vx_2)
 \ppl(\vx_3) \ppl(\vx_4),
 \label{ee4}
 \eea
 where Eq.(\ref{e9}) is used and $K_\pm$ and $\tilde K$ are
 defined as follows:
 \bea
 K_\pm(\vx_1,\cdots,\vx_4)&=&\int d^4 x\ {\cal G}_\pm(x;\vx_1)\cdots{\cal
 G}_\pm(x;\vx_4)\nn\\
 {\tilde K}(\vx_1,\cdots,\vx_4)&=&\frac{1}{4}\sum_p\int d^4x\ {\cal G}_-(x;\vx_{p_1})
 {\cal G}_-(x;\vx_{p_2}) {\cal G}_+(x;\vx_{p_3}){\cal G}_+(x;\vx_{p_4}).
 \label{ee5}
 \eea
 \section{Conclusion}
 Considering the hypersurface t=0 as the boundary (the holographic screen)
 of  Minkowski (Euclidean) space, and using the conformal map between massless
 scalar fields in Euclidean
 (Minkowski) space and Euclidean AdS (the past/future region of dS) we
 obtained the generating function of CFT correlators of operators on the boundary
 of (A)dS space, recognized in
 (A)dS/CFT dual to  free field theories, by solving the corresponding
 action in terms of the Cauchy data given at the boundary. In the
 case of four dimensional $\p^4$-model, we derived the generating function
 up to first order in $g$, the coupling constant.
 We also obtained some exact solutions of the non-linear
 Klein-Gordon equation for $\phi^3$, $\phi^4$ and $\phi^6$-models in $D=6,4,3$
 Euclidean/Minkowski (AdS/dS)  spaces.
 \par
 The conformal map between (Euclidean) Minkowski and (Euclidean
 anti) de Sitter spaces is a useful method to examine (A)dS/CFT for
 various interacting theories, at least at tree-level (quantum corrections may introduce
 a mass scale to the theory). For example, in the case of free spinors, the same
 method affirmed the surface term recognized in AdS/CFT \cite{Hen} to be added to the
 standard Dirac action in the case of dS/CFT \cite{Spin}. Another attempt could be studying
 the $D=4$ minimal-coupled spinor-scalar theory (${\bar \psi}\phi\psi$-model) to
 obtain and verify the role of the surface term considered in \cite{Fat} to
 study AdS/CFT for interacting spinor-scalars action in the case of de Sitter
 space.
 \section*{Acknowledgement} The financial support of Isfahan
 University of Technology is acknowledged.
 \newpage
 \section{Appendix A}
 In this appendix we briefly review free scalar field theory in $D=d+1$ dimensional
 curved spacetime.  The action for the scalar field $\phi$ is
 \be
 S=\int d^Dx\;
 \sqrt{\left|g\right|}\frac{1}{2}\left(g^{\mu\nu}\partial_\mu\phi\partial_\nu\phi -(m^2+\xi
 R)\phi^2\right),
 \ee
 for which the equation of motion is
 \be
 \left(\Box + m^2+\xi R\right)\phi=0,\hspace{1cm}
 \Box=|g|^{-1/2}\partial_\mu\left|g\right|^{1/2}g^{\mu\nu}\partial_\nu.
 \ee
 (With $\hbar$ explicit, the mass $m$ should be replaced by
 $m/\hbar$.) The case with $m=0$ and $\xi=\frac{d-1}{4d}$ is referred to
 as conformal coupling  \cite{Ted}.
 \par
 Using Eq.(\ref{b1}) it is easy to show that the Ricci scalar $R$ for $\mbox{dS}_{d+1}$
 space is $R=d(d+1)$ where we have set the dS radius $\ell=1$. Therefore, the action for
 conformally coupled scalars in $\mbox{dS}_{d+1}$ is
 \be
 S=\int d^Dx\;
 \sqrt{\left|g\right|}\frac{1}{2}\left(g^{\mu\nu}\partial_\mu\phi\partial_\nu\phi -
 (\frac{d^2-1}{4})\phi^2\right),
 \ee
 Similar result can be obtained for AdS space using Eq.(\ref{met}).
 \section{Appendix B}
 In this appendix, we modify the equation of motion of the phi-fourth model in the presence of the brane located at $t=0$,
 and give a method that can be used to calculate the solution of
 equation of motion to arbitrary order of perturbation in $g$,
 the coupling constant. At the first order in $g$, we show that the generating function for the correlators of dual operators on the brane is the one
 given in  Eq.(\ref{b6}).

 Considering a free field theory in the presence of the brane at
 $t=0$, one can simply verify that Eq.(\ref{a42}) gives the correct
 solution to the equation of motion for $t>0$. But  ${\cal G}$ defined in
 Eq.(\ref{g}) not only is a solution of the equation of motion for $t>0$,
 but also indicates the brane located at $t=0$. To see this, we make
 an analytic continuation to complex $k^0$-plane to rewrite
 $\cal G$ as follows,
 \bea
 {\cal G}(\vec x,t;\vec y)&=&\int\frac{d^dk}{(2\pi)^d} e^{-\omega_k
 t}e^{i\vec k.(\vec x-\vec y)}\nn\\
 &=&-2\int\frac{d^{d+1}k}{(2\pi)^{d+1}} \frac{\omega_k}{k^2}e^{i
 k.(x-y)},
 \label{ap-b1}
 \eea
 where $\omega_k=\left|\vec k\right|$, $k^2=(k^0)^2+\omega_k^2$,
 $x^0=t$ and $y^0=0$.
 From Eq.(\ref{ap-b1}) one verifies that
 \bea
 \Box{\cal G}(x;\vec y)&=&2\delta(t)\int\frac{d^dk}{(2\pi)^d}e^{i\vec k.(\vec x-\vec
 y)}\nn\\
 &=&2\delta(t)F(\left|\vec x-\vec y\right|),
 \label{ap-b2}
 \eea
 in which $\Box=\partial_t^2+\partial_{\vec x}^2$ and
 $F(\left|z\right|)$ has been defined in Eq.(\ref{a5-1}). Furthermore,
 \be
 {\cal G}(\vec x,0;\vec y)=\delta^d(\vec x-\vec y).
 \label{rrr}\ee
 Thus
 $\phi$ is Eq.(\ref{a42}) satisfies the modified equation of motion
 given as follows,
 \be
 \Box\phi=2\delta(t)\int d^dyF(\left|\vec x-\vec
 y\right|)\phi_0(\vec y).
 \label{ap-b3}
 \ee
 This is the true equation of motion as it indicates the {\em source} on
 the barne at $t=0$.

 An identity useful to calculate the action $I[\phi]$ given in
 Eq.(\ref{intbypart}) is
 \be
 \partial_t{\cal G}(\vec x,t;\vec y)|_{t=0}=-F(\left|\vec x-\vec
 y\right|),
 \label{ap-b4}
 \ee
 which can be used to show that,
 \be
 I[\phi]=-\frac{1}{2}\int d^dx d^dy\phi_0(\vec x)F(\left|\vec x-\vec
 y\right|)\phi_0(\vec y).
 \label{ap-b5}
 \ee
 The minus sign in the above equation is crucial as it makes the
 action $I[\phi]$ positive valued since for example for $d=3$,
 \be
 F(\left|\vec z\right|)=-\frac{1}{\pi^2}\left|\vec z\right|^{-4}.
 \ee

 In the case of $\phi^4$ model, the equation of motion for $t>0$  is
 \be
 \Box\phi+g\phi^3=0.
 \label{ap-b6}
 \ee
 A solution of  this equation with the boundary condition
 $\phi_0(\vec x)$ given at $t=0$ is Eq.(\ref{b3}),
 \be
 \phi(x)=\int d^d y {\cal G}(x;\vec y)\phi_0(\vec y)-g\int_{0\le y^0\le t}d^{d+1}y G_{\rm
 E}(x,y)\phi^3(y),
 \label{ap-b7}
 \ee
 where
 \bea
 G_{\rm E}(x,y)=-\int\frac{d^{d+1}k}{(2\pi)^{d+1}} \frac{e^{i
 k.(x-y)}}{k^2}.
 \label{ap-b8}
 \eea
 is the Euclidean Green function given in
 Eq.(\ref{Green}).
 In fact from the definition (\ref{ap-b8}) one can show that
 \be
 G_{\rm E}(x,y)=-\int\frac{d^dk}{(2\pi)^d}e^{i\vec
 k.(\vec x-\vec y)}\int
 \frac{dk^0}{2\pi}\frac{e^{ik^0(x^0-y^0)}}{(k^0)^2+\omega_k^2},
 \label{ap-b9}\ee
 and
 \be
 \int\frac{dk^0}{2\pi}\frac{e^{ik^0(x^0-y^0)}}{(k^0)^2+\omega_k^2}=\left\{\begin{array}{lll}-\frac{1}{2\omega_k}e^{-\omega_k(x^0-y^0)}&&x^0>y^0,\\\\
 \frac{1}{2\omega_k}e^{\omega_k(x^0-y^0)}&&x^0<y^0.\end{array}\right.
 \label{ap-b10}\ee
 Eq.(\ref{rrr}) can be used to show that $\phi$ given in Eq.(\ref{ap-b7}) satisfies
 the boundary condition $\phi(\vec x,0)=\phi_0(\vec x)$.
 Furthermore, using the identity $\Box G_{\rm E}=\delta^{d+1}(x-y)$, one
 obtains,
 \be
 \Box \phi+g\phi^3=2\delta(t)\int d^dyF(\left|\vec x-\vec
 y\right|)\phi_0(\vec y),
 \label{ap-b11}\ee
 which is the equation of motion modified by the brane at $t=0$.

 To obtain the action $I[\phi]$ we use Eq.(\ref{b1}). Defining
 \be
 \eta_0(x)=\int d^d y {\cal G}(x;\vec y)\phi_0(\vec y),
 \label{ap-b12}\ee
 It is not difficult to verify that $I[\phi]$ given in Eq.(\ref{b1})
 is equivalent to
 \be
 I[\phi]=\frac{1}{2}\int_{R^3}\eta_0\partial_t\eta_0+\frac{g}{4}\int_{R^4}\phi^4.
 \label{ap-b13}
 \ee
 In fact to obtain the above equation one should only be careful
 about $\delta(t)$ appearing in $\Box \phi$ and the condition
 $0\le y^0\le t$ in the second term in Eq.(\ref{ap-b7}). Thus to the
 lowest order in $g$ the action functional $I[\phi]$ is given by
 Eq.(\ref{b6}). Unfortunately Eq.(\ref{important}) is not correct as we
 will show in the following. In fact there is a first order correction term in $g$ which makes contribution to the higher order terms in $I[\phi]$ and
 will not change the result given in Eq.(\ref{b6}).

 Before going to the second part of this appendix where a method to calculate $\phi$ in terms of
 a  perturbation series in $g$ is given I do like to comment on the claim (\ref{claim}). It is now clear to me that the conditions on
 $K_{\rm E}$ enumerated in section 3,
 can not uniquely determine $K_{\rm E}$ as there are some other functions which satisfy all these conditions.\footnote{I
 am grateful to Karl-Henning Rehren who notified me of this fact on Jan 2005.}
 Unfortunately I have not been able to calculate $K_{\rm E}$ given in Eq.(\ref{b11}) as an explicit function of the coordinates $\vec x_i$'s yet and leave it
 as an open problem.

\begin{figure}[t]
 \centerline{\epsfxsize=2in\epsffile{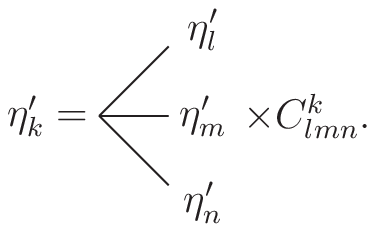}} \caption{Diagrammatic representation of Eq.(\ref{ap-b15}).}
 \label{fig1}
 \end{figure}

\begin{figure}[t]
 \centerline{\epsfxsize=2.3in\epsffile{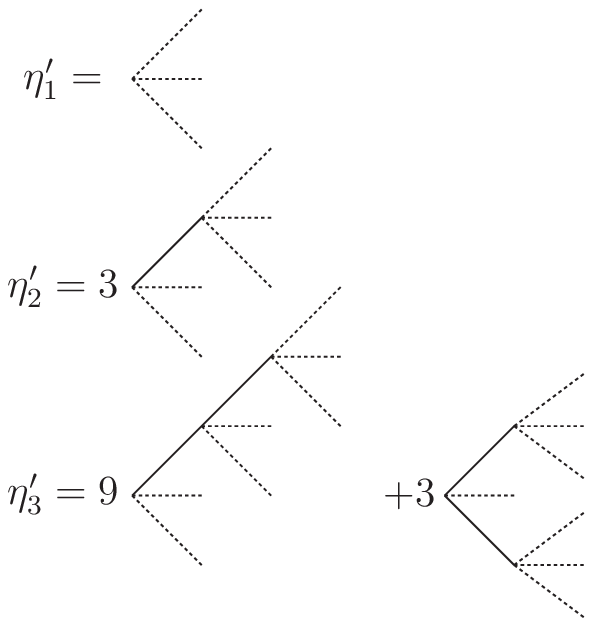}} \caption{Diagrammatic representation of $\eta'_1,\eta'_2,\eta'_3$.}
 \label{fig2}
 \end{figure}

 To solve Eq.(\ref{ap-b7}) we assume that $\phi$ is given as the following series in
 $g$,
 \be
 \phi(x)=\sum_{k=0}g^k\eta_k,
 \label{ap-b14}\ee
 and calculate the functions $\eta_k$. Using
 Eq.(\ref{ap-b7}) it is not difficult to show that,
 \be
 \eta_k(x)=-\int_{0\le y^0\le t}d^{d+1}y G_{\rm
 E}(x,y)\sum_{l+m+n=k-1}(\eta_l\eta_m\eta_n)(y),\hspace{1cm}k\ge1,
 \label{ap-b15}\ee
 where $\eta_0$ is given in Eq.(\ref{ap-b12}). It is easy to use
 Eq.(\ref{ap-b15}) to calculate $\eta_k$. For example,
 \bea
 \eta_1&=&-\int' d^{d+1}y G_{\rm E}(x,y)\eta_0^3(y)\nn\\
 \eta_2&=&-\int' d^{d+1}y G_{\rm E}(x,y)(3\eta_1\eta_0^2)(y)\nn\\&=&3\int' d^{d+1}y G_{\rm
 E}(x,y)\eta_0^2(g)\int' d^{d+1}z G_{\rm E}(y,z)\eta_0^3(z).
 \eea
 In general $\eta_k$ is given by the diagram given in Fig.(\ref{fig1}),
 where $\eta'$ is defined by the identity,
 \be
 \eta(x)=-\int'd^dy G_E(x,y)\eta'(y).
 \ee
 Each line in Fig.(\ref{fig1}) stands for the operation
 \be
 -\int'G_{\rm E},
 \ee
 and the combinatoric factor is
 \be
 C^k_{lmn}=\left\{\begin{array}{lll}1,&&l=m=n,\\3,&&l=m\neq
 n,\\6,&&{\mbox{otherwise}}.\end{array}\right.
 \ee
 More explicitly,
 \be
 C^k_{lmn}=\frac{3!}{S},
 \ee
 in which $S$ is the symmetry factor. For example,
 \bea
 C^1_{000}&=&\frac{3!}{3!},\nn\\
 C^2_{100}&=&\frac{(3!)^2}{2!3!}=3.
 \eea
 In Fig.(\ref{fig2}) we represent $\eta'_k$ for $k=1,2,3$
 diagrammatically. A dashed-line stands for $\eta_0$.

 
\end{document}